\documentclass[draft]{svmult-turin}

\usepackage{type1cm} 

\usepackage{makeidx}         % allows index generation
\usepackage{graphicx}        % standard LaTeX graphics tool
\usepackage[bottom]{footmisc}% places footnotes at page bottom

\usepackage{newtxtext}      % 
\usepackage{newtxmath}      % selects Times Roman as basic font

\newcommand{\fa}{\forall}

\renewcommand{\d}{\partial}
\newcommand{\ubr}{\underbrace}
\newcommand{\lbetr}{\left\lvert}
\newcommand{\rbetr}{\right\lvert}
\newcommand{\abs}[1]{\lbetr {#1}\rbetr}
\newcommand{\lnorm}{\left\lVert}
\newcommand{\rnorm}{\right\lVert}
\newcommand{\norm}[1]{\lnorm {#1}\rnorm}

\renewcommand{\l}{\ensuremath{\left}}
\renewcommand{\r}{\ensuremath{\right}}
\newcommand{\then}{\ensuremath{\Rightarrow}}

\newcommand{\Hp}{\ensuremath{\mathcal{H}}}
\newcommand{\Sp}{\ensuremath{\mathcal{S}}}
\newcommand{\Lp}{\ensuremath{\mathcal{L}}}
\newcommand{\rn}{\ensuremath{\mathbb{R}}}
\newcommand{\cn}{\ensuremath{\mathbb{C}}}
\newcommand{\nn}{\ensuremath{\mathbb{N}}}
\newcommand{\zn}{\ensuremath{\mathbb{Z}}}
\newcommand{\Gf}{\ensuremath{\mathfrak{G}}}

\newcommand{\Af}{\ensuremath{\mathfrak{A}}}
\newcommand{\Bf}{\ensuremath{\mathfrak{B}}}

\renewcommand{\phi}{\varphi}

\renewcommand{\rho}{\varrho}

\renewcommand{\theta}{\vartheta}
\newcommand{\eps}{\varepsilon}

\newcommand{\bra}[1]{\ensuremath{\left\langle{#1}\right|}}
\newcommand{\ket}[1]{\ensuremath{\left|{#1}\right\rangle}}
\newcommand{\tr}{\ensuremath{\operatorname{tr}}}
\newcommand{\id}{\ensuremath{\mathrm{id}}}
\newcommand{\diag}{\ensuremath{\mathrm{diag}}}
\newcommand{\vol}{\ensuremath{\mathrm{vol}}}
\newcommand{\Texp}{\ensuremath{\mathrm{Texp}}}

\allowdisplaybreaks

\begin{document}

\title*{Integrating Gauge Fields in the $\zeta$-formulation of Feynman's path integral}
\titlerunning{Integrating gauge field in the $\zeta$-path integral}
% Use \titlerunning{<name>} for an abbreviated version of
% your contribution title if the original one is too long

\author{Tobias Hartung and Karl Jansen}
% Use \authorrunning{Short Title} for an abbreviated version of
% your contribution title if the original one is too long
\institute{Tobias Hartung \at Department of Mathematics, King's College London, Strand, London WC2R 2LS, United Kingdom \email{tobias.hartung@kcl.ac.uk}
\and Karl Jansen \at NIC, DESY Zeuthen, Platanenallee 6, 15738 Zeuthen, Germany \email{karl.jansen@desy.de}}
%
% Use the package "url.sty" to avoid
% problems with special characters
% used in your e-mail or web address
%
\maketitle

\abstract{
  In recent work by the authors, a connection between Feynman's path integral and Fourier integral operator $\zeta$-functions has been established as a means of regularizing the vacuum expectation values in quantum field theories. However, most explicit examples using this regularization technique to date, do not consider gauge fields in detail. Here, we address this gap by looking at some well-known physical examples of quantum fields from the Fourier integral operator $\zeta$-function point of view. 
}


\section{Introduction}
Feynman's path integral is a fundamental building block of modern quantum field theory. For instance, the time evolution semigroup $(U(t,s))_{t,s\in\rn_{\ge0}}$ of a quantum field theory is a semigroup of integral operators whose kernels are given by the path integral. In terms of the Hamiltonian $H$ of a given quantum field theory, $U$ is the semigroup generated by $-\frac{i}{\hbar}H$, i.e., $U$ formally satisfies $U(t,s)=\Texp\l(-\frac{i}{\hbar}\int_s^t H(\tau)d\tau\r)$ where $\Texp$ is the time-ordered exponential for unbounded operators as to be understood in terms of the time-dependent Hille-Yosida Theorem~(e.g., Theorem 5.3.1 in~\cite{pazy}). Furthermore, the path integral is intimately connected to vacuum expectation values which play two very crucial roles. On one hand, vacuum expectation values are physical and allow for experimental verification and thus to test theories. On the other hand, vacuum expectation values of $n$ field operators (so called $n$-point functions) uniquely determine the quantum field theory by Wightman's Reconstruction Theorem (Theorem 3-7 in~\cite{streater-wightman}).

Let us consider a quantum field theory with Hilbert space $\Hp$ and time evolution semigroup~$U$. Then, the vacuum expectation value $\langle A\rangle$ of an observable $A$ can be expressed as
\begin{align*}\tag{$*$}\label{eq:vev-formal}
  \langle A\rangle=\lim_{T\to\infty+i0^+}\frac{\tr\l(U(T,0)A\r)}{\tr U(T,0)}
\end{align*}
where the denominator $\tr U(T,0)$ is also known as the partition function. Upon closer inspection however, \eqref{eq:vev-formal} reveals one of the major mathematical obstacles. The traces on the right hand side of~\eqref{eq:vev-formal} should be the canonical trace on trace-class operators $\Sp_1(\Hp)$ but for a continuum theory $U(T,0)$ is a bounded, non-compact operator and $U(T,0)A$ is in general an unbounded operator on $\Hp$.

Vacuum expectation values are thus only generally understood in terms of discretized quantum field theories. This is the starting point of lattice quantum field theory for instance and great computational effort is necessary to extrapolate the continuum limit from these discretized vacuum expectation values. If we wish to understand~\eqref{eq:vev-formal} in the continuum however, the traces need to be constructed in such a way that they coincide with the canonical trace on $\Sp_1(\Hp)$ provided $U(T,0),U(T,0)A\in\Sp_1(\Hp)$.

One such trace construction technique are operator $\zeta$-functions. They were introduced by Ray and Singer~\cite{ray,ray-singer} for pseudo-differential operators and first proposed as a regularization method for path integrals in perturbation theory by Hawking~\cite{hawking}. The Fourier integral operator $\zeta$-function approach generalizes the pseudo-differential framework to non-perturbative settings with general metrics (Euclidean and Lorentzian) and includes special cases like Lattice discretizations in a Lorentzian background.

Given an operator $A$ and a trace $\tau$ for which we want to define $\tau(A)$, we construct a holomorphic family $\Af$ such that $\Af(0)=A$ and there exists a maximal open and connected subset $\Omega$ of $\cn$ for which $\Af$ maps $\Omega$ into the domain of $\tau$. In general, we construct $\Af$ such that $\Omega$ contains a half-space $\cn_{\Re(\cdot)<R}:=\{z\in\cn;\ \Re(z)<R\}$ for some $R\in\rn$. Then, we define the $\zeta$-function $\zeta(\Af)$ to be the meromorphic extension of $\Omega\ni z\mapsto\tau(\Af(z))\in\cn$ to an open, connected neighborhood of $0$ (provided it exists). If $\zeta(\Af)$ is holomorphic in a neighborhood of $0$ and $\zeta(\Af)(0)$ depends only on $A$ and not the explicit choice of $\Af$ (that is, if $\Bf$ is another admissible choice of holomorphic family with $\Bf(0)=\Af(0)$, then $\zeta(\Af)(0)=\zeta(\Bf)(0)$), then we can define $\tau(A)$ as $\zeta(\Af)(0)$.

For example, if $A$ is a positive operator whose spectrum $\sigma(A)$ is discrete and free from accumulation points, then we could define $\Af(z):=A^z$ and $\zeta(\Af)$ is given by the meromorphic extension of $z\mapsto\sum_{\lambda\in\sigma(A)\setminus\{0\}}\lambda^z$ (counting multiplicities); hence, giving rise to the name ``operator $\zeta$-function.'' This is precisely how Hawking~\cite{hawking} employed $\zeta$-regularization, it has been used successfully in many physical settings (e.g., the Casimir effect, defining one-loop functional determinants, the stress-energy tensor, conformal field theory, and string theory~\cite{beneventano-santangelo,blau-visser-wipf,bordag-elizalde-kirsten,bytsenko-et-al,culumovic-et-al,dowker-critchley,elizalde2001,elizalde,elizalde-et-al,elizalde-vanzo-zerbini,fermi-pizzocchero,hawking,iso-murayama,marcolli-connes,mckeon-sherry,moretti97,moretti99,moretti00,moretti11,robles,shiekh,tong-strings}), and is related to Hadamard parametrix renormalization~\cite{hack-moretti}. This approach has been fundamental for many subsequent developments as it allows for an effective Lagrangian to be defined~\cite{blau-visser-wipf} as well as heat kernel coefficients to easily be computed~\cite{bordag-elizalde-kirsten}, and implies non-trivial extensions of the Chowla-Selberg formula~\cite{elizalde2001}. Furthermore, the residues have been studied extensively because they give rise to the multiplicative anomaly which appears in perturbation theory~\cite{elizalde-vanzo-zerbini} and contributes a substantial part to the energy momentum tensor of a black hole for instance~\cite{hawking}.

Kontsevich and Vishik~\cite{kontsevich-vishik,kontsevich-vishik-geometry} showed that this construction gives rise to a well-defined (unbounded) trace for pseudo-differential operators. Their approach was later extended to Fourier integral operators~\cite{hartung-phd,hartung-scott}. Since Radzikowski~\cite{radzikowski92,radzikowski96} showed that the operators $U(T,0)A$ and $U(T,0)$ are pseudo-differential operators (Euclidean spacetimes) or more generally Fourier integral operators (Lorentzian spacetimes), we can apply this framework of operator $\zeta$-functions to the definition of vacuum expectation values as it was first done in~\cite{hartung,hartung-iwota} and define a $\zeta$-regularized vacuum expectation value of $A$ to be
\begin{align*}
  \langle A\rangle_\zeta:=\lim_{z\to0}\lim_{T\to\infty+i0^+}\frac{\zeta(U(T,0)\Gf A)}{\zeta(U(T,0)\Gf)}(z)
\end{align*}
where $\Gf$ is a suitable family of Fourier integral operators (usually pseudo-differential) with $\Gf(0)=1$ such that $U(T,0)\Gf A$ and $U(T,0)\Gf$ satisfy the assumptions on the construction of the corresponding operator $\zeta$-functions.

If we consider $\l(U(t,s)\Gf(z)\r)_{s,t\in\rn_{\ge0}}$ to be the time evolution semigroup of a quantum field theory $QFT(z)$, then this essentially means that we construct a ``holomorphic family of quantum field theories $QFT$'' such that the vacuum expectation value of $A$ in $QFT(z)$ is well-defined in Feynman's sense for $z$ in some open subset $\Omega$ of $\cn$ and the vacuum expectation value of $A$ in the quantum field theory $QFT(0)$, that we wish to study, is defined via analytic continuation.

Furthermore, it was recently shown~\cite{hartung-jansen} that this construction of $\zeta$-regularized vacuum expectation values can be understood in terms of a continuum limit of discretized quantum field theories which is accessible using quantum computing. This discretization can be constructed directly in the continuum on general metrics, including Riemannian and Lorentzian spacetimes. Alternatively, the discretization can be constructed from spacetime lattices. Given the universal applicability result~\cite{hartung-jansen} of the Fourier integral operator $\zeta$-function approach to $\zeta$-regularized vacuum expectation values, many examples have been considered in this framework on a mathematically fundamental level~\cite{hartung,hartung-iwota,hartung-jansen}. However, applications of $\zeta$-regularization in the physical literature~\cite{beneventano-santangelo,blau-visser-wipf,bordag-elizalde-kirsten,bytsenko-et-al,culumovic-et-al,dowker-critchley,elizalde2001,elizalde,elizalde-et-al,elizalde-vanzo-zerbini,fermi-pizzocchero,hawking,iso-murayama,marcolli-connes,mckeon-sherry,moretti97,moretti99,moretti00,moretti11,robles,shiekh,tong-strings} have focused on different aspects which leaves a wide gap to demonstrate the practicability of treating quantum field theories with the Fourier integral operator $\zeta$-function regularization in a non-perturbative fashion.

We therefore want to start filling this gap with some fundamental examples of quantum fields which are underlying many gauge field theories. In particular, we will consider free real and complex scalar quantum fields (Sections~\ref{sec:free-real-scalar} and~\ref{sec:free-complex-scalar} respectively) and the free Dirac field (Section~\ref{sec:free-dirac}). Finally, we will consider light coupled to a fermion (Section~\ref{sec:fermion-light}) where we ignore self-interaction of the radiation field for simplicity (the free radiation field has already been discussed in~\cite{hartung-jansen}). The example of light coupling to matter is of particular interest as it is one of the well-known examples of $\zeta$-regularization from the physical literature~\cite{iso-murayama} which we can now understand in terms of the Fourier integral operator approach to $\zeta$-regularized vacuum expectation values.

\section{The free real scalar quantum field}\label{sec:free-real-scalar}
The first example we would like to consider is the free scalar quantum field in $1+1$ dimensions. Its Lagrangian density is given by
\begin{align*}
  \Lp=\frac12(\d_0\phi)^2-\frac12(\d_1\phi)^2.
\end{align*}
Hence, the generalized momentum is
\begin{align*}
  \hat\phi=\d_{\d_0\phi}\Lp=\d_0\phi
\end{align*}
and thus we obtain the Hamiltonian density
\begin{align*}
  h=\hat\phi\d_0\phi-\Lp=\frac12\hat\phi^2+\frac12(\d_1\phi)^2.
\end{align*}
Considering the spatial torus $\rn/X\zn$, the momenta of the quantum field take values in $\frac{2\pi}{X}\zn\setminus\{0\}$ and the dispersion relation $E_p^2=p^2$ yields the energy $E_p=\abs p$ of a particle with momentum $p\in\frac{2\pi}{X}\zn\setminus\{0\}$. Hence, using the canonical quantization of free fields (cf. e.g.~\cite{tong} chapter 2) we obtain the quantized field $\Phi$ and momentum $\Pi$
\begin{align*}
  \Phi(x)=&\sum_{p\in\frac{2\pi}{X}\zn\setminus\{0\}}\frac{1}{\sqrt{2XE_p}}\l(a_pe^{ipx}+a_p^\dagger e^{-ipx}\r)\\
  \Pi(x)=&\sum_{p\in\frac{2\pi}{X}\zn\setminus\{0\}}(-i)\sqrt{\frac{E_p}{2X}}\l(a_pe^{ipx}-a_p^\dagger e^{-ipx}\r)
\end{align*}
where $a_p$ and $a_p^\dagger$ are the normalized annihilation and creation operators for a particle of momentum $p$. In other words, they satisfy the canonical commutation relations $[a_p,a_q]=[a_p^\dagger,a_q^\dagger]=0$ and $[a_p,a_q^\dagger]=\delta_{p,q}$. Plugging these expressions into the Hamiltonian density ($\phi\rightsquigarrow\Phi$ and $\hat\phi\rightsquigarrow\Pi$) and integrating over $\rn/X\zn$ then yields the Hamiltonian
\begin{align*}
  H=&\frac12\sum_{p,q\in\frac{2\pi}{X}\zn\setminus\{0\}}\Bigg(\frac{-\sqrt{E_pE_q}}{2}\l(a_pa_q\delta_{p,-q}-a_pa_q^\dagger\delta_{p,q}-a_p^\dagger a_q\delta_{p,q}+a_p^\dagger a_q^\dagger\delta_{p,-q}\r)\\
  &+\frac{1}{2\sqrt{E_pE_q}}\l(-pqa_pa_q\delta_{p,-q}+pqa_pa_q^\dagger\delta_{p,q}+pqa_p^\dagger a_q\delta_{p,q}-pqa_p^\dagger a_q^\dagger\delta_{p,-q}\r)\Bigg)\\
  =&\frac12\sum_{p\in\frac{2\pi}{X}\zn\setminus\{0\}}\frac{1}{2E_p}\l((-E_p^2+p^2)(a_pa_{-p}+a_p^\dagger a_{-p}^\dagger)+(E_p^2+p^2)(a_pa_p^\dagger+a_p^\dagger a_p)\r)\\
  =&\frac12\sum_{p\in\frac{2\pi}{X}\zn\setminus\{0\}}E_p(2a_p^\dagger a_p+1)\\
  =&\sum_{p\in\frac{2\pi}{X}\zn\setminus\{0\}}E_p\l(a_p^\dagger a_p+\frac12\r)
\end{align*}
since $a_pa_p^\dagger=a_p^\dagger a_p+1$ and $E_p^2=p^2$. Here, the term $\sum_{p\in\frac{2\pi}{X}\zn\setminus\{0\}}E_pa_p^\dagger a_p$ is precisely what we expect to see since $a_p^\dagger a_p$ counts the number of particles with momentum~$p$. The term $\sum_{p\in\frac{2\pi}{X}\zn\setminus\{0\}}\frac{E_p}{2}$ on the other hand diverges. In the physics literature, you usually encounter a renormalization argument at this point or the Hamiltonian is directly redefined to be normally ordered, and the term is dropped. Therefore, we define the normally ordered Hamiltonian to be
\begin{align*}
  H_n:=\sum_{p\in\frac{2\pi}{X}\zn}E_pa_p^\dagger a_p
\end{align*}
where we artificially added the $p=0$ term which corresponds to the ``there are no particles'' case.

On the other hand, we are looking to use a $\zeta$-regularized framework and this additional term 
\begin{align*}
  \sum_{p\in\frac{2\pi}{X}\zn\setminus\{0\}}\frac{E_p}{2}=&\frac{2\pi}{X}\sum_{k\in\nn}k\text{ ``$=$'' }\frac{2\pi}{X}\zeta_R(-1)=-\frac{\pi}{6X}
\end{align*}
can be interpreted as such where $\zeta_R$ denotes the Riemann $\zeta$-function. In other words, we can define a $\zeta$-regularized Hamiltonian
\begin{align*}
  H_\zeta:=H_n-\frac{\pi}{6X}.
\end{align*}
It is interesting to note that $H_\zeta$ and $H_n$ coincide in the limit $X\to\infty$ which eventually we need to perform if we want to obtain vacuum expectation values in $1+1$~Minkowski space. However, physically this constant has no impact at all since it is not an observable. This relies on the fact that we cannot measure ``absolute'' energies but only differences in energy. The choice between $H_\zeta$ and $H_n$ is therefore similar to the choice between measuring temperature in Kelvin~($H_n$) or Celsius~($H_\zeta$).

In order to use the $\zeta$-formalism, we need to find Fourier integral operator representations of $H_\zeta$ and $H_n$. Since the two only differ by a constant, we will only consider $H_n$ for the moment. Let $H_n^1$ be the restriction of $H_n$ to the space generated by at most single particle states. Calling the vacuum state $\ket0$, we can obtain all single particle states $\ket p=a_p^\dagger\ket0$ using the corresponding creation operator and, since $a_q^\dagger a_q\ket p=\delta_{p,q}\ket p$, we directly obtain $H_n^1\ket p=\abs p\ket p$. In other words, $\l(\ket{p}\r)_{p\in\frac{2\pi}{X}\zn}$ is an orthonormal basis of the Hilbert space spanned by all at most single particle states which we can thus identify with $\ell_2\l(\frac{2\pi}{X}\zn\r)$ and therefore with $L_2(\rn/X\zn)$ as well. In particular, we have the correspondence
\begin{align*}
  \ell_2\l(\frac{2\pi}{X}\zn\r)\ni\ket p\longleftrightarrow\l(x\mapsto e^{ipx}\r)\in L_2(\rn/X\zn)
\end{align*}
and obtain the $L_2(\rn/X\zn)$ representation
\begin{align*}
  H_n^1=\abs\d.
\end{align*}
In order to allow multiple particles to exist, suppose we have the $N$~particle state $\ket P=\ket{P_0,P_1,\ldots,P_{N-1}}=\l(\prod_{j=0}^{N-1}a_{P_j}^\dagger\r)\ket0$. This state can be represented as a sum of permutations of tensor products $\ket P=\frac{1}{N!}\sum_{\pi\in S_N}\bigotimes_{j=0}^{N-1}\ket{P_{\pi(j)}}$, where $S_N$ denotes the symmetric group on the set $N$, in the symmetric tensor product $S\bigotimes_{j=1}^{N}\l(L_2(\rn/X\zn)\ominus\cn\r)$. The Hilbert space $\Hp$ is then the Fock space given by the Hilbert space completion of $\bigoplus_{N\in\nn_0}S\bigotimes_{j=1}^N\l(L_2(\rn/X\zn)\ominus\cn\r)$.

Thus, states in $\Hp$ are of the form
\begin{align*}
  \ket\Psi =& a_0\ket0\oplus\bigoplus_{N\in\nn}\sum_{j_0^N,\ldots,j_{N-1}^N}a_{j_0^N,\ldots,j_{N-1}^N}\ket{p_{j_0^N},\ldots,p_{j_{N-1}^N}}\\
  \ket\Phi =& b_0\ket0\oplus\bigoplus_{N\in\nn}\sum_{k_0^N,\ldots,k_{N-1}^N}b_{k_0^N,\ldots,k_{N-1}^N}\ket{p_{k_0^N},\ldots,p_{k_{N-1}^N}}
\end{align*}
and the inner product is given by
\begin{align*}
  \langle\Psi,\Phi\rangle=a_0^*b_0+\sum_{n\in\nn}\sum_{j_0^N,\ldots,j_{N-1}^N}\sum_{k_0^N,\ldots,k_{N-1}^N}a_{j_0^N,\ldots,j_{N-1}^N}^*b_{k_0^N,\ldots,k_{N-1}^N}\prod_{m=0}^{N-1}\langle p_{j_m^N},p_{k_m^N}\rangle.
\end{align*}

Given a pure $N$~particle state, we deduce that the restriction $H_n^N$ of $H_n$ to the $N$~particle Hilbert space $S\bigotimes_{j=1}^{N}\l(L_2(\rn/X\zn)\ominus\cn\r)$ is given by
\begin{align*}
  H_n^N=\sum_{j=1}^N\l(\bigotimes_{k=1}^{j-1}\id\r)\otimes H_n^1\otimes\l(\bigotimes_{k=j+1}^{N}\id\r)=\sum_{j=1}^N\frac{2\pi}{X}\abs{\d_j}
\end{align*}
and, finally, we can represent $H_n$ on the Fock space $\Hp$ as
\begin{align*}
  H_n=\bigoplus_{N\in\nn}H_n^N=\diag\l(\l(\sum_{j=1}^N\frac{2\pi}{X}\abs{\d_j}\r)_{N\in\nn_0}\r).
\end{align*}
In this case, the energy of the state $\ket\Psi$ is given by
\begin{align*}
  \langle\Psi,H_n\Psi\rangle_{\Hp}=&\sum_{N\in\nn}\sum_{j_0^N,\ldots,j_{N-1}^N}\abs{a_{j_0^N,\ldots,j_{N-1}^N}}^2\sum_{k=0}^{N-1}E_{p_{j_k^N}}
\end{align*}
i.e., precisely the expression we were looking for. In particular, this expression is minimal if and only if each summand is zero which implies $\ket\Psi=\ket0$. In other words, the vacuum expectation of $H_n$ is 
\begin{align*}
  \langle H_n\rangle=\bra0H_n\ket0=0.
\end{align*}
Of course, this directly implies
\begin{align*}
  \langle H_\zeta\rangle=\bra0H_n\ket0-\frac{\pi}{6X}\langle0|0\rangle=-\frac{\pi}{6X}
\end{align*}
as expected.

\subsection{The $\zeta$-regularized vacuum expectation values of $H_n$ and $H_\zeta$}

Let us now compare the true vacuum expectations $\langle H_n\rangle=0$ and $\langle H_\zeta\rangle=-\frac{\pi}{6X}$ to the $\zeta$-regularized vacuum expectation values $\langle H_n\rangle_\zeta$ and $\langle H_\zeta\rangle_\zeta$. Again, we will start with $H_n$. However, if we try to na\"ively ignore that we have a Fock space here,
\begin{align*}
  &\langle H_n\rangle_\zeta\\
  =&\lim_{z\to0}\lim_{T\to\infty}\frac{\int_{\bigtimes_{N\in\nn}\rn^N}\sum_{N\in\nn}e^{i\frac{2\pi T}{X}\sum_{n=1}^N\norm{\xi_{N,n}}}\frac{2\pi}{X}\sum_{n=1}^N\norm{\xi_{N,n}}\prod_{m=1}^N\norm{\xi_{N,m}}^zd\xi}{\int_{\bigtimes_{N\in\nn}\rn^N}\sum_{N\in\nn}e^{i\frac{2\pi T}{X}\sum_{n=1}^N\norm{\xi_{N,n}}}\prod_{m=1}^N\norm{\xi_{N,m}}^zd\xi}\\
  =&\lim_{z\to0}\lim_{T\to\infty}\frac{\sum_{N\in\nn}\frac{2\pi}{X}\sum_{n=1}^N\prod_{m=1}^N\int_{\rn^N}e^{i\frac{2\pi T}{X}\norm{\xi}}\norm{\xi}^{z+\delta_{m,n}}d\xi}{\sum_{N\in\nn}\sum_{n=1}^N\prod_{m=1}^N\int_{\rn^N}e^{i\frac{2\pi T}{X}\norm{\xi}}\norm{\xi}^{z}d\xi}\\
  =&\lim_{z\to0}\lim_{T\to\infty}\frac{\sum_{N\in\nn}\frac{2\pi N}{X}\int_{\rn^N}e^{i\frac{2\pi T}{X}\norm{\xi}}\norm{\xi}^{z+1}d\xi\l(\int_{\rn^N}e^{i\frac{2\pi T}{X}\norm{\xi}}\norm{\xi}^{z}d\xi\r)^{N-1}}{\sum_{N\in\nn}N\l(\int_{\rn^N}e^{i\frac{2\pi T}{X}\norm{\xi}}\norm{\xi}^{z}d\xi\r)^N}\\
  =&\lim_{z\to0}\lim_{T\to\infty}\frac{\sum_{N\in\nn}\frac{2\pi N\l(\vol\d B_{\rn^N}\r)^N}{X}\int_{\rn_{>0}}e^{i\frac{2\pi T}{X}r}r^{z+N}dr\l(\int_{\rn_{>0}}e^{i\frac{2\pi T}{X}r}r^{z+N-1}dr\r)^{N-1}}{\sum_{N\in\nn}N\l(\vol\d B_{\rn^N}\r)^N\l(\int_{\rn_{>0}}e^{i\frac{2\pi T}{X}r}r^{z+N-1}dr\r)^N}\\
  =&\lim_{z\to0}\lim_{T\to\infty}\frac{\sum_{N\in\nn}N\l(\vol\d B_{\rn^N}\r)^N\Gamma(z+N+1)\Gamma(z+N)^{N-1}\l(i\frac{2\pi T}{X}\r)^{-Nz-N^2-1}}{\sum_{N\in\nn}\frac{2\pi N\l(\vol\d B_{\rn^N}\r)^N}{X}\Gamma(z+N)^N\l(i\frac{2\pi T}{X}\r)^{-Nz-N^2}}
\end{align*}
shows that we have not completely $\zeta$-regularized since the series might not be convergent for sufficiently small $\Re(z)$. Instead we need to introduce a regularization for the summation over $N$ as well. For instance, let 
\begin{align*}
  \alpha_N^z:=\Gamma(z+N+1)^{-1}\Gamma(z+N)^{-N}\l(i\frac{2\pi T}{X}\r)^{Nz}N^z\l(\vol\d B_{\rn^N}\r)^{Nz}.
\end{align*}
Then $\fa N\in\nn:\ \alpha_N^0=1$ and we obtain
\begin{align*}
  &\langle H_n\rangle_\zeta\\
  =&\lim_{z\to0}\lim_{T\to\infty}\frac{\int_{\bigtimes_{N\in\nn}\rn^N}\sum_{N\in\nn}\alpha_N^ze^{i\frac{2\pi T}{X}\sum_{n=1}^N\norm{\xi_{N,n}}}\frac{2\pi}{X}\sum_{n=1}^N\norm{\xi_{N,n}}\prod_{m=1}^N\norm{\xi_{N,m}}^zd\xi}{\int_{\bigtimes_{N\in\nn}\rn^N}\sum_{N\in\nn}\alpha_N^ze^{i\frac{2\pi T}{X}\sum_{n=1}^N\norm{\xi_{N,n}}}\prod_{m=1}^N\norm{\xi_{N,m}}^zd\xi}\\
  =&\lim_{z\to0}\lim_{T\to\infty}\ubr{\frac{\sum_{N\in\nn}\frac{2\pi N^{1+z}\l(\vol\d B_{\rn^N}\r)^{N(1+z)}}{X}\Gamma(z+N)^{-1}\l(i\frac{2\pi T}{X}\r)^{-N^2-1}}{\sum_{N\in\nn}N^{1+z}\l(\vol\d B_{\rn^N}\r)^{N(1+z)}\Gamma(z+N+1)^{-1}\l(i\frac{2\pi T}{X}\r)^{-N^2}}}_{\in O\l(\frac{1}{T}\r)}\\
  =&0
\end{align*}
which coincides with the $\langle H_n\rangle$.

Regarding $H_\zeta$, let $\Gf$ be a gauged Fourier integral operator such that $\Gf(0)=1$. Then, $e^{-\frac{iT\pi}{6X}}e^{iTH_n}\Gf(0)=e^{iTH_\zeta}$ and
\begin{align*}
  \langle H_\zeta\rangle_\zeta=&\lim_{z\to0}\lim_{T\to\infty}\frac{\zeta\l(e^{-\frac{iT\pi}{6X}}e^{iTH_n}\Gf H_\zeta\r)}{\zeta\l(e^{-\frac{iT\pi}{6X}}e^{iTH_n}\Gf\r)}(z)\\
  =&\lim_{z\to0}\lim_{T\to\infty}\frac{e^{-\frac{iT\pi}{6X}}\l(\zeta\l(e^{iTH_n}\Gf H_n\r)-\frac{\pi}{6X}\zeta\l(e^{iTH_n}\Gf\r)\r)}{e^{-\frac{iT\pi}{6X}}\zeta\l(e^{iTH_n}\Gf\r)}(z)\\
  =&\langle H_n\rangle_\zeta-\frac{\pi}{6X}
\end{align*}
implies $\langle H_\zeta\rangle_\zeta=\langle H_\zeta\rangle=-\frac{\pi}{6X}$.

\subsection{The $N\to\infty$ particle limit}
Alternatively, we can consider the Hamiltonian 
\begin{align*}
  H_n^{\le N}=\sum_{j=1}^N\abs{\d_j}
\end{align*}
in the ``up to $N$ particle Hilbert space'' $L_2((\rn/X\zn)^N)$ where $\ket{P_0,\ldots,P_{k-1}}$ is embedded as $\bigotimes_{j\in k}\ket{P_j}\otimes\bigotimes_{j\in N-k}\ket0$. Physically, taking the limit $N\to\infty$ says that we are only considering states that have finitely many particles. The $\zeta$-regularized vacuum energy is then computed as
\begin{align*}
  &\lim_{N\to\infty}\langle H_n^{\le N}\rangle_\zeta\\
  =&\lim_{N\to\infty}\lim_{z\to0}\lim_{T\to\infty}\frac{\int_{\rn^N}e^{i\frac{2\pi T}{X}\sum_{n=1}^N\norm{\xi_{n}}}\frac{2\pi}{X}\sum_{n=1}^N\norm{\xi_{n}}\prod_{n=1}^N\norm{\xi_{n}}^zd\xi}{\int_{\rn^N}e^{i\frac{2\pi T}{X}\sum_{n=1}^N\norm{\xi_{n}}}\prod_{n=1}^N\norm{\xi_{n}}^zd\xi}\\
  =&\lim_{N\to\infty}\lim_{z\to0}\lim_{T\to\infty}\frac{\frac{2\pi}{X}\sum_{n=1}^N\prod_{m=1}^N\int_{\rn^N}e^{i\frac{2\pi T}{X}\norm{\xi}}\norm{\xi}^{z+\delta_{m,n}}d\xi}{\sum_{n=1}^N\prod_{m=1}^N\int_{\rn^N}e^{i\frac{2\pi T}{X}\norm{\xi}}\norm{\xi}^{z}d\xi}\\
  =&\lim_{N\to\infty}\lim_{z\to0}\lim_{T\to\infty}\frac{\frac{2\pi N}{X}\int_{\rn^N}e^{i\frac{2\pi T}{X}\norm{\xi}}\norm{\xi}^{z+1}d\xi\l(\int_{\rn^N}e^{i\frac{2\pi T}{X}\norm{\xi}}\norm{\xi}^{z}d\xi\r)^{N-1}}{N\l(\int_{\rn^N}e^{i\frac{2\pi T}{X}\norm{\xi}}\norm{\xi}^{z}d\xi\r)^N}\\
  =&\lim_{N\to\infty}\lim_{z\to0}\lim_{T\to\infty}\frac{\frac{2\pi}{X}\int_{\rn_{>0}}e^{i\frac{2\pi T}{X}r}r^{z+N}dr\l(\int_{\rn_{>0}}e^{i\frac{2\pi T}{X}r}r^{z+N-1}dr\r)^{N-1}}{\l(\int_{\rn_{>0}}e^{i\frac{2\pi T}{X}r}r^{z+N-1}dr\r)^N}\\
  =&\lim_{N\to\infty}\lim_{z\to0}\lim_{T\to\infty}\frac{\frac{2\pi}{X}\Gamma(z+N+1)\Gamma(z+N)^{N-1}\l(i\frac{2\pi T}{X}\r)^{-Nz-N^2-1}}{\Gamma(z+N)^N\l(i\frac{2\pi T}{X}\r)^{-Nz-N^2}}\\
  =&0
\end{align*}
in this setting.

\section{Free complex scalar quantum fields}\label{sec:free-complex-scalar}
Complex scalar fields are generalizations of real scalar fields which allow for the creation of antiparticles. More precisely, in a real scalar field the particle is its own antiparticle. The distinction between particles and antiparticles for the complex scalar field becomes obvious once they are quantized. Writing a complex scalar field $\psi=\frac{\phi_1+i\phi_2}{\sqrt2}$ as the sum of two real scalar fields $\phi_1$ and $\phi_2$ with creation operators $b^\dagger$ and $c^\dagger$, and expanding the field operator as a sum of planar waves yields
\begin{align*}
  \Psi(x)=&\sum_{p\in M}\frac{1}{\sqrt{2XE_p}}\l(b_pe^{ipx}+c_p^\dagger e^{-ipx}\r)\\
  \Psi^\dagger(x)=&\sum_{p\in M}\frac{1}{\sqrt{2XE_p}}\l(b_p^\dagger e^{-ipx}+c_p e^{ipx}\r)
\end{align*}
on $\rn/X\zn$ where $M=\frac{2\pi}{X}\zn\setminus\{0\}$ is the set of momenta. This furthermore implies the conjugate momentum
\begin{align*}
  \Pi(x)=&\sum_{p\in M}i\sqrt{\frac{E_p}{2X^N}}\l(b_p^\dagger e^{-ipx}-c_pe^{ipx}\r)\\
  \Pi^\dagger(x)=&\sum_{p\in M}(-i)\sqrt{\frac{E_p}{2X^N}}\l(b_pe^{ipx}-c_p^\dagger e^{-ipx}\r).
\end{align*}
If we consider the charge operator $Q=i\int\Pi(x)\Psi(x)-\Psi^*(x)\Pi^*(x)dx$ we directly obtain
\begin{align*}
  Q=\sum_{p\in M}c_pc_p^\dagger-b_p^\dagger b_p
\end{align*}
which is not normally ordered. The normally ordered charge $Q_n$ is thus given by 
\begin{align*}
  Q_n=\sum_{p\in M}c_p^\dagger c_p-b_p^\dagger b_p.
\end{align*}
This again can be explained using a $\zeta$-argument and the commutator relation $[c_p,c_p^\dagger]=1$. More precisely, we need to $\zeta$-regularize the series $\sum_{p\in M}1$ which is nothing other than $\tr\id$ on $\rn/X\zn$. Since $\id$ has no critical degree of homogeneity, $\tr\id:=\zeta(z\mapsto\abs\nabla^z)(0)$ exists and is a well-defined constant (in fact, it is $2\zeta_R\l(0\r)=-1$ where $\zeta_R$ is the Riemann $\zeta$-function), i.e.,
\begin{align*}
  Q_\zeta=Q_n+\tr\id=Q_n-1.
\end{align*}

As for the Hamiltonian, we repeat the same calculation we did in the real case but with Lagrangian $\d^\mu\phi^*\d_\mu\phi$ instead of $\d^\mu\phi\d_\mu\phi$ and obtain the normally ordered Hamiltonian
\begin{align*}
  H_n=\sum_{p\in\frac{2\pi}{X}\zn}E_p(b_p^\dagger b_p-c_p^\dagger c_p)
\end{align*}
which differs from the $\zeta$-regularized Hamiltonian by a constant again. This also shows the interesting effect that antiparticles appear with negative energy in the theory which allows us to reproduce the Feynman-St\"uckelberg interpretation of antiparticles. Considering the wave propagator under time-reversal $\exp(itH_n)\rightsquigarrow\exp(-itH_n)$ we obtain an algebraically equivalent theory with reversed roles for $b_p$ and $c_p$. In other words, antiparticles are particles that move backwards in time and creation and annihilation of particle-antiparticle pairs can be seen as a particle reversing the direction it travels through time.

In any case, the negative energies yield the up to $N$ particle and $N$ anti-particle Hamiltonian
\begin{align*}
  H_n^{\le N}=\sum_{j=1}^N\l(\abs{\d_{1,j}}-\abs{\d_{2,j}}\r)
\end{align*}
on $L_2((\rn/X\zn)^{2N})$ which directly implies that $\lim_{N\to\infty}\langle H_n^{\le N}\rangle_\zeta=\langle H_n\rangle_\zeta=0$ since the degrees of homogeneity are identical to the ones in the real scalar field case.

\section{The Dirac field}\label{sec:free-dirac}
The free Dirac field is closely related to the complex scalar field but we are now considering spinor valued fields, assume that the creation and annihilation operators satisfy the canonical anticommutator relations, and possibly introduce a mass term~$m$. Hence, our fields on the spatial torus $(\rn/X\zn)^N$ are
\begin{align*}
  \Psi(x)=&\sum_{p\in M}\sum_{s\in S}\frac{1}{\sqrt{2X^NE_p}}\l(b_p^su_p^se^{ipx}+c_p^{s\dagger}v_p^s e^{-ipx}\r)\\
  \Psi^\dagger(x)=&\sum_{p\in M}\sum_{s\in S}\frac{1}{\sqrt{2X^NE_p}}\l(b_p^{s\dagger}u_p^{s\dagger} e^{-ipx}+c_pv_p^{s\dagger} e^{ipx}\r)
\end{align*}
where $S$ is the set of spins, $M$ the set of momenta, and $u$ and $v$ are spinors, i.e., they satisfy
\begin{enumerate}
\item[(i)] $(\gamma^\mu p_\mu-m)u^s_p=0$
\item[(ii)] $(\gamma^\mu p_\mu+m)v^s_p=0$
\item[(iii)] $u_p^{r\dagger}u_p^s=v_p^{r\dagger}v_p^s=2E_p\delta^{rs}$
\item[(iv)] $u_p^{r\dagger}v_{-p}^s=v_p^{r\dagger}u_{-p}^s=0$
\end{enumerate}
where $(p^\mu)_{\mu}=(E_p,p)^T$, $E_p=\sqrt{\langle p,p\rangle+m^2}$, and the $\gamma$-matrices are given in the Dirac basis $\gamma^0=\begin{pmatrix}1&0\\0&-1\end{pmatrix}$ and $\gamma^k=\begin{pmatrix}0&\sigma^k\\-\sigma^k&0\end{pmatrix}$ with the Pauli matrices $\sigma^1=\begin{pmatrix}0&1\\1&0\end{pmatrix}$, $\sigma^2=\begin{pmatrix}0&-i\\i&0\end{pmatrix}$, and $\sigma^3=\begin{pmatrix}1&0\\0&-1\end{pmatrix}$. Plugging everything into the Dirac Hamiltonian density $\Psi^\dagger\gamma^0(-i\gamma^j\d_j+m)\Psi$ and integrating then yields
\begin{align*}
  H=\sum_{p\in M}\sum_{s\in S}E_p(b_p^{s\dagger} b_p^s+c_p^s c_p^{s\dagger})
\end{align*}
and $c_p^s c_p^{s\dagger}=1-c_p^{s\dagger} c_p^{s}$ yields the normally ordered Hamiltonian
\begin{align*}
  H_n=\sum_{p\in M}\sum_{s\in S}E_p(b_p^{s\dagger} b_p^s-c_p^{s\dagger}c_p^s).
\end{align*}
For $m=0$ this is precisely the same situation we had for the complex scalar field just with an additional summation over spins.

For $m>0$ we still have the question whether we can normally order the Hamiltonian using a $\zeta$-argument again. In other words, we need to $\zeta$-regularize the trace of an operator with kernel $\sqrt{\norm{\xi}^2+m^2}$ but for $\norm\xi>m$ we observe the asymptotic expansion
\begin{align*}
  \sqrt{\norm{\xi}^2+m^2}=\sum_{j\in\nn_0}{\frac{1}{2}\choose j}\norm\xi^{1-2j}m^{2j}
\end{align*}
which has a degree of homogeneity $-N$ if and only if $N$ is odd. In particular, the residue trace is given by ${\frac{1}{2}\choose \frac{N+1}{2}}m^{N+1}\vol\d B_{\rn^N}$. Hence, $\zeta$-regularization fails to normally order this Hamiltonian.

However, this is no problem in the light of vacuum expectation values as we are taking quotients of $\zeta$-functions. Hence, the presence of poles simply means that the value of $\frac{\zeta(U\Gf A)}{\zeta(U\Gf)}(0)$ is given by the quotient of residues rather than the quotient of constant Laurent coefficients.

\section{Coupling a fermion of mass $m$ to light in $1+1$ dimensions}\label{sec:fermion-light}
Coupling light to matter in $1+1$ dimensions is one of the text-book examples of $\zeta$-regularization in the physical literature because it is a toy model for QED. In particular, the Schwinger model which has $m=0$ has been studied extensively (cf. e.g.~\cite{iso-murayama}). Here, we will show how the well-known applications of $\zeta$-regularization tie into the framework of $\zeta$-regularized vacuum expectation values as discussed in~\cite{hartung,hartung-iwota,hartung-jansen}.

In order to consider coupling a fermion to a gauge field $(A_\mu)_\mu$, we will restrict our considerations to a fermion in $1+1$ dimensions with a constant background field. This ignores the self-interaction of the gauge field which gives an additional term to the Hamiltonian that has already been discussed in~\cite{hartung-jansen}.

In the present case, and using the temporal gauge $A_0=0$, $A:=A_1$, the (fermionic coupling) Hamiltonian on $\rn/X\zn$ is given by
\begin{align*}
  H_F=\int_0^X\Psi(x)^\dagger\l((i\d-eA)\sigma_3+m\r)\Psi(x)dx
\end{align*}
where $e$ is the coupling constant, $\sigma_3=\begin{pmatrix}1&0\\0&-1\end{pmatrix}$, and $\Psi$ is the spinor field which we endow with anti-periodic boundary conditions $\Psi(x+X)=-\Psi(x)$ (this is allowed because $\Psi$ is an auxiliary field; all physical quantities are composed of sesquilinear forms in $\Psi$ which are periodic).

To study this system, we will first expand $\Psi$ into eigenmodes of $(i\d-eA)\sigma_3+m$, i.e., we are looking to solve
\begin{align*}
  (i\d-eA)\sigma_3
  \begin{pmatrix}
    \psi^+\\0
  \end{pmatrix}
  =&\eps^+-m
  \begin{pmatrix}
    \psi^+\\0
  \end{pmatrix}\quad\text{and}\quad
  (i\d-eA)\sigma_3
  \begin{pmatrix}
    0\\\psi^-
  \end{pmatrix}
  =-\eps^--m
  \begin{pmatrix}
    0\\\psi^-
  \end{pmatrix}.
\end{align*}
These imply
\begin{align*}
  \psi^\pm(x)=\frac{1}{\sqrt{X}}\exp\l(-ie\int_0^xA(y)dy-i(\eps^\pm\mp m) x\r)
\end{align*}
where 
\begin{align*}
  e^{-i\pi-2i\pi n}\Psi(x)=-\Psi(x)=\Psi(x+X)
\end{align*}
implies that $\eps^\pm$ has to satisfy
\begin{align*}
  &-i\pi-2i\pi n-ie\int_0^xA(y)dy-i(\eps^\pm\mp m) x\\
  =&-ie\int_0^xA(y)dy-i(\eps^\pm\mp m) x-ie\int_0^XA(y)dy-i(\eps^\pm\mp m) X.
\end{align*}
In other words, the eigenvalues are given by
\begin{align*}
  \fa n\in\zn:\ \eps^\pm_n:=\frac{\pi}{X}+\frac{2\pi}{X} n-\frac{e\oint A}{X}\pm m=\frac{2\pi}{X}\l(n+\frac{1}{2}\pm mX-\frac{e\oint A}{2\pi}\r)
\end{align*}
where $\oint A:=\int_0^XA(y)dy$. For brevity, we will write $C^\pm:=\frac{e\oint A}{2\pi}\mp mX$.

First quantization of $\Psi$, then introduces annihilation operators $a_n$ and $b_n$ for the upper and lower components of $\Psi$ with $\{a_m,a_n^\dagger\}=\{b_m,b_n^\dagger\}=\delta_{m,n}$ and $\Psi$ is given by
\begin{align*}
  \Psi(x)=\sum_{n\in\zn}
  \begin{pmatrix}
    \psi_n^+(x)a_n\\
    \psi_n^-(x)b_n
  \end{pmatrix}
  =\frac{1}{\sqrt{X}}\sum_{n\in\zn}
  \begin{pmatrix}
    \exp\l(-ie\int_0^xA(y)dy-i(\eps_n^+-m) x\r)a_n\\
    \exp\l(-ie\int_0^xA(y)dy-i(\eps_n^-+m) x\r)b_n
  \end{pmatrix}.
\end{align*}
In particular, this implies
\begin{align*}
  H_F=&\int_0^X\Psi(x)^\dagger\l((i\d-eA)\sigma_3+m\r)\Psi(x)dx=\sum_{n\in\zn}(\eps_n^+a_n^\dagger a_n-\eps_n^-b_n^\dagger b_n).
\end{align*}
At this point, we will split our considerations into the positive ($a_n$) and negative ($b_n$) chirality sectors. The positive sector has the Hamiltonian $H_+:=\sum_{n\in\zn}\eps_n^+a_n^\dagger a_n$ and chiral charge $Q_+:=\sum_{n\in\zn}a_n^\dagger a_n$. Since there is no minimum energy, we define the $N^+$-vacuum of the positive chirality sector by filling all states with energies $\eps_n^+$ where $n<N^+$. 

To compute the $N^+$-chiral charge $\langle Q_+\rangle_{N^+}=\sum_{n\in\zn_{<N^+}}1$ and $N^+$-vacuum energy $\langle H_+\rangle_{N^+}=\sum_{n\in\zn_{<N^+}}\eps_n^+$, we use $\zeta$-regularization. The gauge family $\Gf_+(z):=\abs{H_+}^z$ makes the computation easily accessible on the spectral side. With this choice of gauge, we observe 
\begin{align*}
  \langle Q_+\rangle_{N^+,\zeta}=&\lim_{z\to0}\sum_{n\in\zn_{<N^+}}\abs{\eps_n^+}^z\\
  =&\lim_{z\to0}\sum_{n\in\zn_{<N^+}}\abs{n+\frac{1}{2}-C^+}^z\\
  =&\lim_{z\to0}\sum_{n\in\nn_0}\abs{N^+-n-\frac{1}{2}-C^+}^z\\
  =&\zeta_H\l(0;\frac{1}{2}+C^+-N^+\r)
\end{align*}
where $\zeta_H$ is the Hurwitz $\zeta$-function (analytically continued in the second argument as well). Using ther Bernoulli polynomials $B_n$ - which are defined as $B_0:=1$, $B_n'=nB_{n-1}$, and $n\ge1\ \then\ \int_0^1B_n(x)dx=0$ - non-positive integer values of $\zeta_H$ are given by $\zeta_H(-n;x)=-\frac{B_{n+1}(x)}{n+1}$. In particular, we will need $\zeta_H(0;x)=\frac{1}{2}-x$ and $\zeta_H(-1;x)=-\frac{1}{2}\l(\l(x-\frac{1}{2}\r)^2-\frac{1}{12}\r)$, the former of which directly implies
\begin{align*}
  \langle Q_+\rangle_{N^+,\zeta}=&N^+-C^+.
\end{align*}
Similarly,
\begin{align*}
  \langle H_+\rangle_{N^+,\zeta}=&\lim_{z\to0}\sum_{n\in\zn_{<N^+}}\eps_n^+\abs{\eps_n^+}^z\\
  =&-\frac{2\pi}{X}\zeta_H\l(-1;\frac{1}{2}+C^+-N^+\r)\\
  =&-\frac{\pi}{X}\l(\l(C^+-N^+\r)^2-\frac{1}{12}\r)\\
  =&-\frac{\pi}{X}\l(\langle Q_+\rangle_{N^+,\zeta}^2-\frac{1}{12}\r).
\end{align*}

The negative chirality sector has chiral charge $Q_-:=\sum_{n\in\zn}b_n^\dagger b_n$ and Hamiltonian $H_-:=\sum_{n\in\zn}(-\eps_n^-)b_n^\dagger b_n$. Again, we introduce an $N^-$-vacuum filling all energy states $-\eps_n^-$ with $n\ge N^-$ and choose the gauge family $\Gf_-(z):=\abs{H_-}^z$. This yields 
\begin{align*}
  \langle Q_-\rangle_{N^-,\zeta}=&\lim_{z\to0}\sum_{n\in\zn_{\ge N^-}}\abs{\eps_n^-}^z
  =\zeta_H\l(0;\frac{1}{2}-C^-+N^-\r)
  =C^--N^-
\end{align*}
and
\begin{align*}
  \langle H_-\rangle_{N^-,\zeta}=&\lim_{z\to0}\sum_{n\in\zn_{\ge N^-}}(-\eps_n^-)\abs{\eps_n^-}^z\\
  =&-\frac{2\pi}{X}\zeta_H\l(-1;\frac{1}{2}-C^-+N^-\r)\\
  =&\frac{\pi}{X}\l(\l(N^--C^-\r)^2-\frac{1}{12}\r)\\
  =&\frac{\pi}{X}\l(\langle Q_-\rangle_{N^-,\zeta}^2-\frac{1}{12}\r).
\end{align*}

Combining both sectors then yields the charge $Q:=Q_++Q_-$, the chiral charge $Q_5:=Q_+-Q_-$, their $N^+$-$N^-$-vacuum expectations
\begin{align*}
  \langle Q\rangle_{N^+,N^-,\zeta}=&N^+-C^+-N^-+C^-=N^+-N^-+2mX,\\
  \langle Q_5\rangle_{N^+,N^-,\zeta}=&N^++N^--C^+-C^-=N^++N^--2\frac{e\oint A}{2\pi},
\end{align*}
and the ground state energy of the fermion
\begin{align*}
  \langle H_F\rangle_{N^+,N^-,\zeta}=&\langle H_+\rangle_{N^+,\zeta}+\langle H_-\rangle_{N^-,\zeta}
  =\frac{\pi}{X}\l(\langle Q_+\rangle_{N^+,\zeta}^2+\langle Q_-\rangle_{N^-,\zeta}^2-\frac{1}{6}\r).
\end{align*}
This combined calculation above can be expressed in terms of Fourier integral operator $\zeta$-functions as $\langle\Omega\rangle_\zeta=\lim_{T\to\infty+i0^+}\frac{\zeta(U\Gf\Omega)}{\zeta(U\Gf)}(0)$ where $\Gf(z)=\abs{H_+}^z\oplus\abs{H_-}^z$.

\section{Conclusion}
In this paper we provided a number of fundamental examples using the Fourier integral operator $\zeta$-function regularization for systems that are relevant in high energy physics. We demonstrated analytically that we obtain the correct vacuum expectation values within this framework and directly addressed the non-trivial problem of treating gauge fields using this point of view. In particular, we discussed scalar fields in sections~\ref{sec:free-real-scalar} (real) and~\ref{sec:free-complex-scalar} (complex), and the Dirac field in section~\ref{sec:free-dirac}. Additionally, we have shown in section~\ref{sec:fermion-light} how one of the canonical applications of $\zeta$-regularization in the physics literature (light coupling to a fermion) appears as a special case of the Fourier integral operator $\zeta$-function approach. This opens the door to also study problems where no analytic solution exists and where the $\zeta$-regularization has to be evaluated numerically, e.g. on a quantum computer as demonstrated in~\cite{hartung-jansen}.

\end{document}